\documentclass[twocolumn,reprint,nobibnotes,altaffilletter,amsmath,amssymb,amsfonts]{revtex4-1}
\usepackage[latin1]{inputenc}
\usepackage{graphicx}
\usepackage{amsmath}
\usepackage{latexsym}
\usepackage{epsfig}
\usepackage{color}
\usepackage[caption=false]{subfig}

\begin{document}
\title{Casimir-like forces  at the  percolation transition}

\author{Nicoletta Gnan$^{1,2}$}
\email{nicoletta.gnan@roma1.infn.it}

\author{Emanuela Zaccarelli$^{1,2}$}
\author{Francesco Sciortino$^{2}$}

\affiliation{ $^1$ CNR-ISC, UOS Sapienza, P.le A. Moro2, I-00185, Roma, Italy}
\affiliation{ $^2$  Dipartimento di Fisica, Universit\`a di Roma ``Sapienza'', P.le A. Moro 2, I-00185, Roma, Italy }

\begin{abstract}
Percolation and critical phenomena show common features such as scaling and universality. 
Colloidal particles, immersed in a solvent close to criticality,  experience long-range effective forces,
named critical Casimir forces. 
Building on the analogy between critical phenomena and percolation, we explore the possibility of observing long-range forces near a percolation threshold. To this aim we numerically evaluate  the effective potential between two colloidal particles dispersed in a chemical sol and we show that it becomes attractive and long-ranged on approaching the sol percolation transition.  We develop a theoretical description based on a polydisperse Asakura-Oosawa model  which captures the divergence of the interaction range, allowing us to interpret such effect in terms of  depletion interactions in a structured solvent.  
 Our results provide the geometric analogue of the critical Casimir force,  suggesting  a novel way for tuning colloidal interactions by controlling the clustering properties of the solvent.
\end{abstract}

\maketitle

\section{Introduction}  

Effective interactions play an important role in the physics of colloidal dispersions~\cite{Likos}. A notable
example is provided by 
depletion interactions, i.e. interactions arising  from the presence of a cosolute (e.g. polymers, surfactants) in the suspension. The pioneering works by Asakura-Oosawa (AO) and Vrij~\cite{AO,Vrij} have shown that, when two hard-sphere (HS) colloids are immersed in a solution in the presence of small co-solute particles, the latter are excluded from the available volume between colloids when the two are closer than the cosolute typical size.
As a result, a net entropy-driven depletion attraction arises.
In general, the strength and the range of the depletion attraction can be tuned by modifying the cosolute concentration and  size.

Depletion interactions have thus far mostly been exploited  for monodisperse cosolutes, such as non-adsorbing polymers or   hard spheres, generating short-range effective forces. A solvent composed by cosolutes of
different size  could introduce a structure in the effective potential  controlled by the different co-solute
length scales~\cite{knoben}. Interestingly, if the co-solutes are constituted by a chemical sol (i.e. a sol composed by particles linked into clusters through irreversible bonds) close to its percolation locus, all length scales will enter in $V_{eff}$, but in a scale-free mode.  Indeed, when the percolation threshold is approached from the sol phase, the cluster size distribution follows a universal power-law dependence  and clusters of all sizes are present (up to a cut-off which is function of the distance from percolation).  
Thus it is legitimate to ask whether two colloids, immersed in a sol of clusters close to percolation, experience a long-ranged effective force, whose characteristic length scale diverges at percolation. 
This question becomes even more interesting if  we consider the analogy between percolation and thermal critical phenomena. Percolation theory describes the growth of clusters in a system on approaching the percolation threshold, the point at which an infinite spanning cluster appears~\cite{Stauffer}. Similarly, the theory of critical phenomena describes the growth of correlated regions on approaching a second order critical point, where the size of the correlated regions diverges~\cite{Domb}. Clusters of different sizes in percolation play the  same role as  the thermal critical fluctuations close to the second-order critical point~\cite{Coniglio}, both being described by scale-free distributions, whose first moment shows a power-law behaviour approaching the transition.   Both  the connectivity length in percolation and the correlation length in critical phenomena diverge at the transition. 

It is well-known that colloidal particles  immersed in a solvent which is close to a second-order critical
point experience   long-range effective forces.   These forces originate
from the confinement of the solvent critical fluctuations  between the  surfaces of distinct colloids~\cite{Fisher}.  The  (universal) resulting effective potential 
decays with an exponential law controlled by the thermal correlation length of the solvent, diverging at the critical point~\cite{Fisher,GambassiPRE}.  These forces, named critical Casimir forces for their analogy with
the Casimir effect occurring when the electromagnetic field  is confined between two metal surfaces, have been measured in recent experiments~\cite{Hertlein}.
Due to their universal nature, critical Casimir interactions do not depend on the specific properties of the solvent but only on the geometry of the confining surfaces and on their ability to absorb the solvent (boundary conditions), giving rise to both attractive or repulsive interactions~\cite{BechingerEPL,GnanJCP}, which  have been exploited to induce colloidal aggregation~\cite{Buzzaccaro,  Bonn, GambassiReply, BonnReply,Schall, Bechinger, GnanSM, SchallNat}. It has also been conjectured that proteins in membranes of living cells experience weak long-range critical Casimir forces~\cite{Machta}.

\begin{figure*}[!ht]
\subfloat[]{\includegraphics[width=18cm]{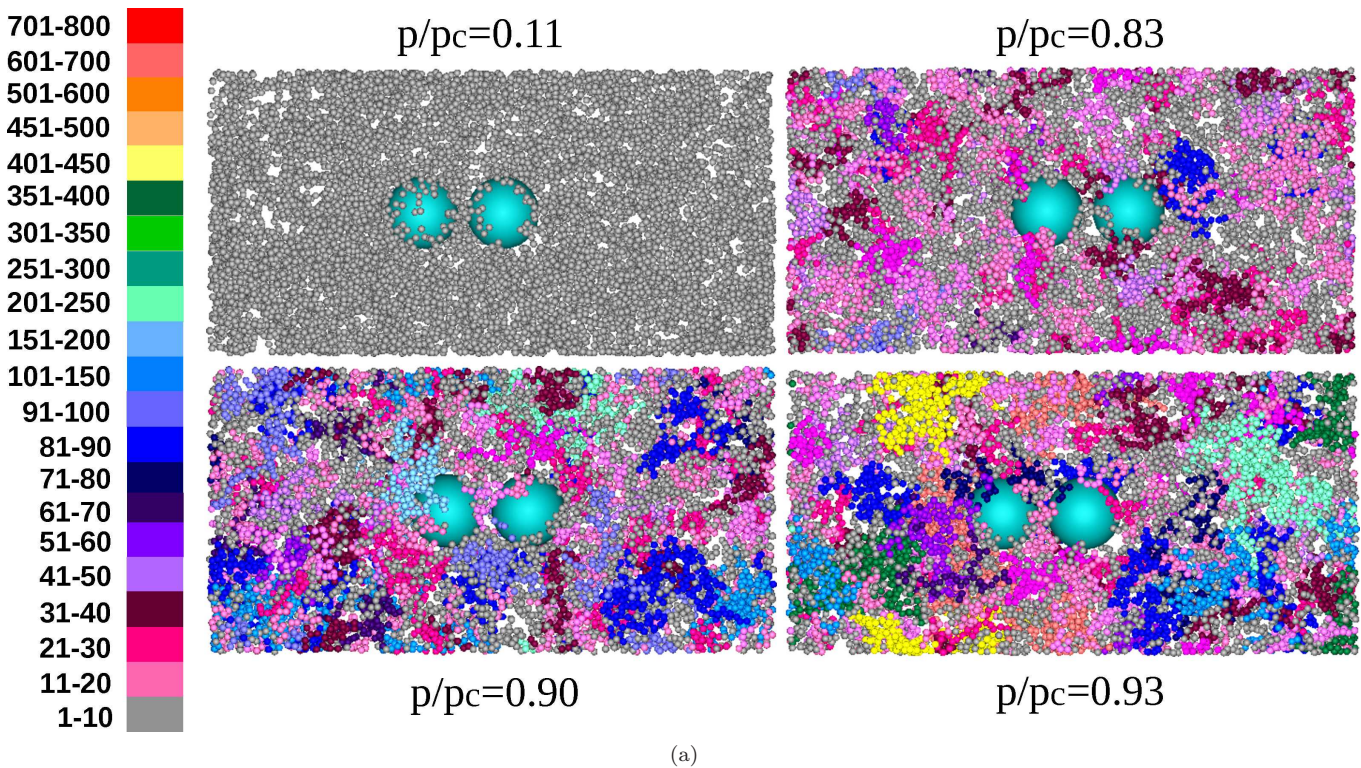}}
    \caption{\textbf{Clustering of the sol close to percolation:} Snapshots of the system for different $p$, with $p\rightarrow p_c$. Each snapshot shows the two colloids immersed in the sol of clusters on approaching the percolation treshold. Clusters of different sizes are represented in a different color (see legend).}\label{fig:figure1}
\end{figure*}

Since percolation shares scaling and universality with critical phenomena, we may expect a mechanism analogous  to the critical Casimir effect to take place  when colloidal particles are immersed in a sol  close to its percolation threshold.  Here we show that indeed a long-range force, created by confining the fluctuations of the cluster sizes,  develops between   colloids and that its interaction range is controlled by  the  sol connectivity length $\xi$.  The resulting effective potentials are compatible with the picture of depletion induced by a polydisperse system. To this aim we perform simulations of two colloidal particles immersed in a chemical sol close to its percolation point.
Fixing the total packing fraction occupied by the clusters, we explore different distances from percolation,
i.e.    cluster size distributions with different cut-off. 
Our results provide evidence that the analogy between percolation and critical phenomena can be exploited to induce novel kind of effective forces between colloidal particles, that are controlled by the clustering properties of the solvent.

\section{Results:}

\subsection{Effective potentials in a sol close to percolation}
We perform Monte Carlo (MC) simulations to evaluate the effective potential $V_{eff}$ between two hard-sphere (HS) colloids of diameter $\sigma_c$ immersed in a fluid composed of clusters. Clusters are made by $N=10836$ hard-sphere monomers of size $\sigma_m = 0.1 \sigma_c$, randomly connected with a maximum functionality of three (as described in the Methods section)  
and are treated as rigid objects, which are allowed to translate and rotate, interacting between themselves and with the two colloids via excluded volume repulsion only.  
Consistent with the hypothesis of  a chemical  sol (irreversible bonds),  clusters  do not break nor coalesce.
The total cluster packing fraction is fixed to $\phi=0.052$ while the distance to the percolation transition (and the associated cluster distributions) changes. 
In simulations, the distance from the transition is controlled by measuring  the fraction $p$ of formed bonds $p$~\cite{Stauffer}. $p$ is the analogous of the temperature in critical phenomena and  its critical values at the percolation threshold  is indicated with $p_c$. 
Snapshots of the system at different $p$ are shown in Fig.~\ref{fig:figure1}.

We have generated, as described in the Methods section, cluster  distributions for  different values of $p$ for $p\rightarrow p_c$. For each selected $p$, we also evaluate the connectivity length $\xi$  defined as~\cite{Stauffer},

\begin{equation}\label{eq:xi2}
 \xi =\left[\frac{2\sum_s R_s^2 s^2 n(s)}{\sum_s s^2 n(s)}\right]^{1/2}
 \end{equation}
\noindent where $n(s)$ is the number of clusters of size $s$ (with the constraint $\sum_s s n(s)=N$), and $R_s$ is the radius of gyration of a cluster  composed of $s$ monomers  with positions $\vec{r}_i$ : $R_s=\left[(1/2s^2)\sum_{ij}|\vec{r}_i-\vec{r}_j|^2\right]^{1/2}$.

In the course of the MC simulation, the volume around the two hard-sphere colloids  is explored by clusters of different size.  Hence, the two colloids experience around them the presence of different clusters that fluctuate in size.  Such fluctuations are analogous to the critical density fluctuations in the critical Casimir effect, and are responsible of the emergence of a long-range force when confined between the surfaces of the two colloids.  
\begin{figure}[!ht]
\includegraphics[width=8.5cm]{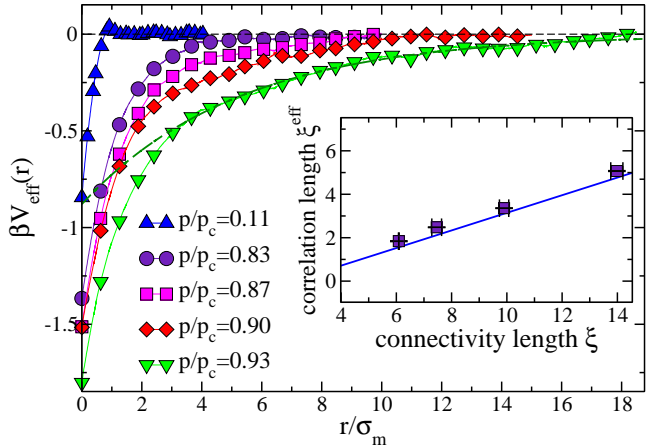}
\caption{\textbf{Effective potentials between two colloids in a sol of clusters:} Evolution of the effective potential $V_{eff}$ on approaching the percolation threshold at fixed sol packing fraction $\phi=0.052$. $r$ is the surface-to-surface distance between the two colloids.
The thick line shows a typical exponential fit to the data in the interval $r/\sigma_m >3$, which is used to estimate the  correlation length $\xi^{eff}$. Inset: correlation length $\xi^{eff}$ extracted from $V_{eff}$ (squares)  against the connectivity length defined in Eq.~\ref{eq:xi2}, both in units of $\sigma_1$. Errors on the $x$ and $y$ axis are respectively the standard error in evaluating the connectivity length over several configurations, and the standard error in performing the exponential fit for extracting the correlation length. The latter is of the order of $1\%$. The solid line is calculated using the theoretical modeling proposed in Eq.~\ref{eq:AO_Result}. The linear relation suggests that  $V_{eff}$ diverges with the same power-law of $\xi$ at the percolation transition. }\label{fig:figure2}
\end{figure}
Figure~\ref{fig:figure2} shows the evolution of the effective potential for different $p$, exploring the range from $p/p_c \approx 0.1$ to $p/p_c \approx 0.93$. For $p/p_c >0.8$ the system is sufficiently close to percolation to sample the universal features of $n(s)$ (as explained in the Methods section). To probe values of  $p/p_c>0.93 $  would require prohibitively larger simulation boxes. 
Far from percolation, 
when the sol is mainly composed of monomers, we recover the 
depletion potential between two colloids immersed in a hard-sphere fluid~\cite{Lajovic, Dijkstra} and
$V_{eff}$ shows a typical oscillatory behaviour, whose characteristic length scale is controlled by the monomer size.  On increasing $p$, $V_{eff}$ turns completely attractive and the  
interaction range becomes longer and longer.  The long distance behaviour is well described by an exponential decay $\exp(-r/\xi^{eff})$, the same functional form that applies to the critical Casimir potential. 
To show that the interaction range is controlled by the connectivity length of the sol, we compare the
correlation length $\xi^{eff}$, extracted from the exponential fit modelling the decay of $V_{eff}$,   
with the connectivity length $\xi$ of the sol  (Eq.~\ref{eq:xi2}) in the inset of Fig.~\ref{fig:figure2}. 
For all investigated $p$ values, we find a linear relation between $\xi$ and $\xi^{eff}$, 
reinforcing the hypothesis that, close to percolation, the range of $V_{eff}$ is indeed related to the typical size of the clusters composing the sol. 

\begin{figure*}[!ht]
\includegraphics[width=18cm]{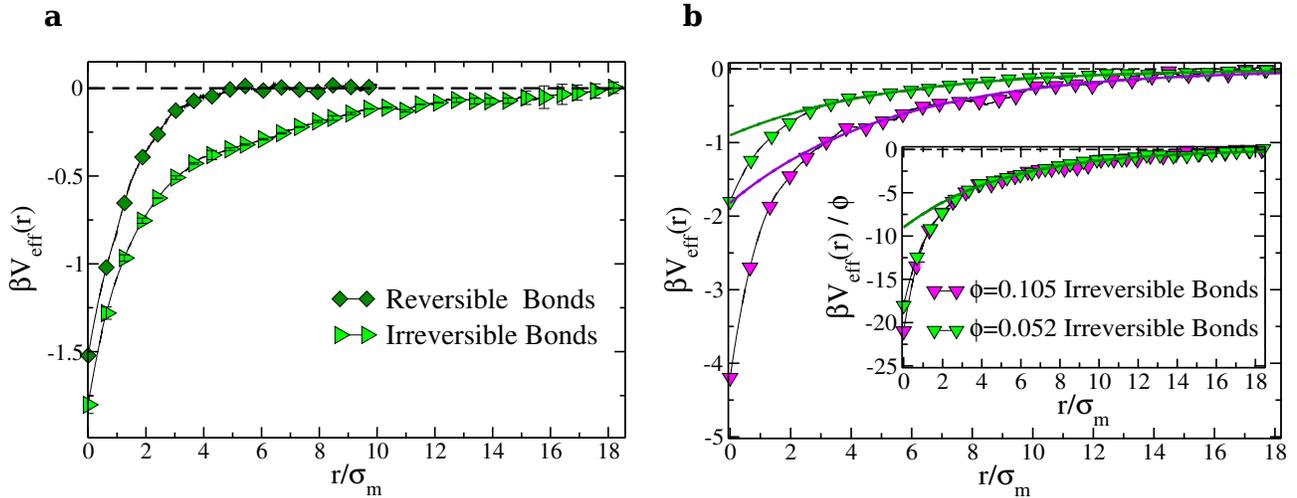}
    \caption{\textbf{Role of the lifetime of sol clusters}  (a) Effective potential between two HS colloids in a solution of particles that forms reversible (diamonds) and irreversible bonds (triangles), with the same cluster size distribution at $p/p_c=0.93$ and $\phi=0.052$. The irreversible potential is averaged over two different realisations of the cluster fluid. Error bars for this potential are calculated from the difference of the two realisations, finding that for $r/\sigma_m <15$ the error does not exceed $5\%$ of the estimated value. (b) Comparison of the  investigated effective potentials generated by irreversible clusters at two different packing fractions along  an  iso-$p$ line (see Methods). Full lines are exponential fits of the long-range tail of $V_{eff}$ . The best fit decay length is  $\approx 5  \sigma_m$ for $\phi=0.052$ and  $\approx 5.3 \sigma_m$ for $\phi=0.105$. The inset shows that the long-distance part of the potentials superimpose within numerical error when the two potentials are scaled by the packing fraction.}\label{fig:figure3}
\end{figure*}
To provide evidence that the long range phenomenon originates from the connectivity properties of the
sol, we calculate --- for $p/p_c \approx 0.93$ ---  the effective potential  for a system 
with the  same cluster size distribution but  in which  bonds are reversible and clusters break and reform in time. This corresponds to select an energy scale for the bond interaction and  a temperature  for which in equilibrium, the same fraction of bonds  among co-solute monomers is present (see Methods).     
A comparison between $V_{eff}(r)$ calculated for reversible and irreversible bonds is  shown in Fig.~\ref{fig:figure3}(a). We notice the dramatic effect of the finite bond lifetime on the range of the effective potential: when clusters are reversible, the potential is attractive only up to a few monomer sizes.  
In order to provide further evidence  that the connectivity length controls $V_{eff}$ at large distances, we show in Fig.~\ref{fig:figure3}(b) the comparison between $V_{eff}$ for $p/p_c=0.93$ calculated respectively at $\phi=0.052$ (as in Fig.~\ref{fig:figure2}) and at $\phi=0.105$. Since $p/p_c$ is the same in the two cases, the sol is characterised by the same cluster distribution (see Methods), and hence by the same $\xi$.   We find that the long-distance part of the potential is well described by an exponential function having the same decay length for both values of the sol packing fraction, supporting the possibility of an universal behaviour at large distances.


\subsection{Generalized Asakura-Oosawa model for a polydisperse depletant}
To gain a deeper insight in the mechanism that controls  the range of $V_{eff}$ close to percolation in a chemical sol we now develop a  theoretical framework based on the analogy with
depletion interactions. When small depletant particles are added to a colloidal suspension, an attractive entropic force between colloids  builds up, due to the exclusion of the
depletant from the region between particles when their relative distance is
comparable or smaller than the depletant diameter~\cite{Likos}.  In the venerable  Asakura-Oosawa-Vrij (AO) model~\cite{AO,Vrij},  introduced to describe the effective interactions between colloids in a solution of non-interacting polymers,  depletants are modelled  as an ideal gas which interacts via hard-core repulsion with the HS colloids only. In the case of monodisperse depletant particles of radius $R$, the AO effective potential between two HS colloids at a surface-to-surface distance $r$ and depletant number density $\rho$ is 

\begin{eqnarray}\label{eq:AO}
\beta V_{AO}(r,R,\rho) &=
-\pi\rho \left(2R-r \right) \left[\left(\frac{R \sigma_c}{2}+\frac{2R^2}{3}\right) \right. \\ \nonumber
& \left. -\frac{r}{2}\left(\frac{\sigma_c}{2}+\frac{R}{3}\right)-\frac{r^2}{12} \right] \Theta(2R-r).
\end{eqnarray}

where the $\Theta$ function indicates that the potential vanishes for distance longer than $2R$.
We propose to model the sol close to percolation as a polydisperse hard-sphere system distributed according to the cluster size distribution $n(s)$. Each cluster  is represented as a sphere of radius $R_s$, the cluster gyration radius,  reducing the problem to that of two colloids immersed in a sea of ideal depletant particles of different size.  
Following AO,  clusters interact only with the colloids via a  hard-core repulsion.
Since  $\sigma_m$ is significantly smaller than  $\sigma_c$, 
only very close to percolation  the largest cluster size becomes comparable to 
the colloid size. Apart from this small region,  the depletion hypothesis is  valid~\cite{Dijkstra2}.
\begin{figure}\label{fig:data}
\begin{center}
\includegraphics[width=8.5cm]{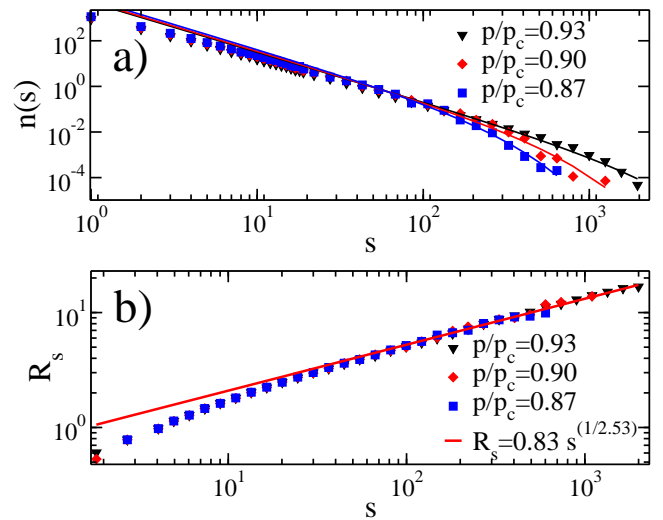}
\caption{\textbf{Clustering properties of the sol:}(a) Cluster size distribution $n(s)$ of the sol employed in MC simulations and (b) radius of gyration $R_s$ as a function of the cluster size $s$ for different $p$, with $p\rightarrow p_c$.
Lines in (a) are power-law fits to $n(s)$ modulated by an exponential function, according to Eq.~\ref{eq:ns}. The solid line in (b) is a fit to $R_s$ following Eq.~\ref{eq:gyration}.
The resulting fit values $s_c$ and $R_1$  are used for evaluating the effective potential in Eq.~\ref{eq:AO_eff}. Note that a theoretical description based on universal properties of cluster fluids close to percolation applies only for cluster sizes above $50$.}
\label{fig:data}
\end{center}
\end{figure}

Close to percolation, the cluster size distribution $n_s$ assumes the universal form~\cite{Stauffer,Tartaglia}
\begin{equation}\label{eq:ns}
n(s)=   \frac{N s^{-\tau} e^{-\frac{s}{s_c}}}{  s_c^{2-\tau} \Gamma(2-\tau,s_c^{-1})  }, 
\end{equation}
where
$\tau=2.18$ is a critical exponent  (in the random percolation universality class) and $s_c$ controls the exponential  cut-off of the power-law distribution,  approaching infinity
at percolation. Moreover $\Gamma(x,y)$ is the incomplete $\Gamma$ function entering via the normalization condition $\int_{1}^{\infty} s n(s)ds=N$.  Summing over all clusters, the resulting potential is $\beta V_{AO}^{eff}(r)=\int_{1}^{\infty}  \beta V_{AO}(r,R_s,n(s)/V) ds$, where $V$ is the volume .

\noindent Building on the universal properties of the clusters shape close to percolation~\cite{Stauffer} it is possible to relate the number of monomers $s$ in the cluster to $R_s$ via the
 fractal exponent $D$, whose universal value (in random percolation theory) is $D=2.53$,
\begin{equation}\label{eq:gyration}
R_s= R_1 s^{1/D}.
\end{equation}
Hence, the total potential becomes,
\begin{eqnarray}
 \label{eq:AO_eff}
&&\beta V_{AO}^{eff}(r)= -\pi    \int_{(\frac{r}{2R_1})^D}^{\infty} ds
\rho_1 \frac{s^{-\tau} e^{-\frac{s}{s_c}}}{  s_c^{2-\tau} \Gamma(2-\tau,s_c^{-1})  }\\ \nonumber
 && \left(2R_s-r \right)  \left[\left(\frac{R_s \sigma_c}{2}+\frac{2R_s^2}{3}\right)
 -\frac{r}{2}\left(\frac{\sigma_c}{2}+\frac{R_s}{3}\right)-\frac{r^2}{12} \right] 
\end{eqnarray}
where $\rho_1$ is the monomer number density and
the lower integration limit $(\frac{r}{2R_1})^D$  in Eq.~\ref{eq:AO_eff} accounts for the $\Theta$ function in each AO contribution. This indicates that only clusters with diameter larger than $r$ participate in building $\beta V_{eff}^{AO}(r)$.   Finally, integrating over the cluster size we obtain,
 \begin{eqnarray}\label{eq:AO_Result}
 \beta V_{AO}^{eff}(r) &= -\frac{\pi \rho_1}{12 s_c \Gamma(2-\tau,s_c^{-1})}\left\lbrace r^2(r+3\sigma_c)\,\Gamma[1 - \tau,\frac{(r/2R_1)^{D}}{s_c}]\right.\nonumber \\
 & \left. -12r\sigma_c R_1 s_c^{1/D}\,\Gamma\left[1 +\frac{1}{D}-\tau,\frac{(r/2R_1)^{D}}{s_c}\right] \right.\nonumber \\
 & \left. -12(r-\sigma_c)(R_1 s_c^{1/D})^2 \,\Gamma\left[1 +\frac{2}{D}-\tau,\frac{(r/2R_1)^{D}}{s_c}\right]\right.\nonumber \\
 & \left. +16 (R_1 s_c^{1/D})^3\,\Gamma\left[1 +\frac{3}{D}-\tau,\frac{(r/2R_1)^{D}}{s_c}\right] \right\rbrace. \nonumber \\
 \end{eqnarray}
\noindent This functional form depends on the percolation exponents and, for large $r$, its asymptotic behaviour is 

\begin{equation}
\beta V^{eff}_{A0}(r\gg R_1)\sim -(r/2R_1)^{(3-\tau D-D)}\exp[-(r/2R_1)^{D}/s_c].
 \end{equation}

To compare the  theoretical predictions based on the AO approach with
the effective potential previously calculated, we first verify that the clusters 
size distribution and the cluster gyration radius employed in the  MC simulations are consistent with the scaling laws predicted by percolation theory 
and extract the corresponding $s_c$ value.  Fig.~\ref{fig:data} shows $n(s)$  and 
$R_s$ with respect to the theoretical predictions of Eq.~\ref{eq:ns} and Eq.~\ref{eq:gyration}. 
For clusters larger than $50$ monomers,  the model-independent scaling laws properly describe the data, suggesting that for $r>2R_{s=50}$,  theoretical predictions 
can be meaningfully compared with the effective potentials calculated from MC simulations.

\begin{figure}
\begin{center}
\includegraphics[width=8.5cm]{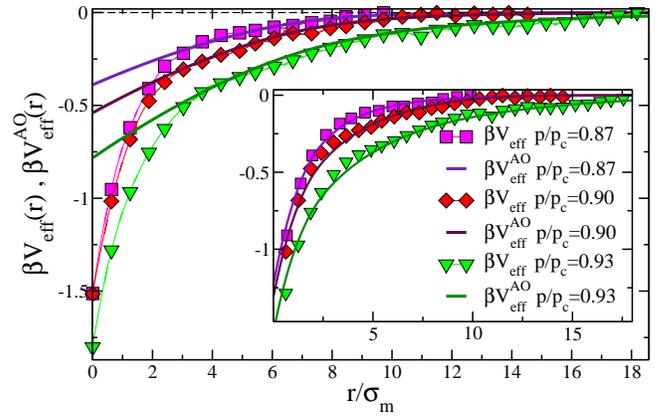}
\caption{\textbf{Comparison between simulations and theory:} symbols correspond to $V_{eff}$ extracted from simulations while solid lines are $V_{AO}^{eff}$ calculated from Eq.~\ref{eq:AO_eff}.  Since universal features of the pre-percolating fluid appear only for clusters larger than $50$ monomers (see Fig.~\ref{fig:data}), theoretical and numerical results show a good agreement only for distances $r> 2R_{s=50}$. Inset: comparison between $V_{eff}$ (symbols) extracted from simulations and $V_{AO}^{eff}$ (solid lines) obtained by repeating the calculation of the effective potential of Eq.~\ref{eq:AO_eff}, in which the exact cluster size configurations employed in simulations are used. Including  the gyration radius  of each cluster in the expression of $V_{AO}^{eff}$ allows us to compare theoretical and numerical results also in the region $r< 2R_{s=50}$. }
\label{fig:V_AO}
\end{center}
\end{figure}
The resulting  $V_{AO}^{eff}$ potentials for different  $p$ are reported  in Fig.~\ref{fig:V_AO}. A surprisingly good agreement between MC results and the theoretical model is found   for $r > 2R_{s=50}\sim 7.8\sigma_m$, confirming that the effect of the sol can be modelled as a depletion interaction acting on all length scales associated with the clusters. We also extract a characteristic decay length of $V_{eff}^{AO}(r)$, employing an exponential fit similarly to what done for the numerical MC data. This can be used to build a relation with $\xi$,  yielding the curve reported in the inset of Fig.~\ref{fig:figure2}, which
closely follows the MC simulation results.  The self-similar nature of the cluster size distribution and its widening on approaching  percolation do control the interaction range.

To strengthen even more the comparison with $V_{AO}^{eff}$, we have repeated the evaluation of the
effective potential   $V_{AO}^{eff}$ in Eq.~\ref{eq:AO_eff}, by numerically summing over the very same cluster configurations employed in simulations, associating each cluster with its own gyration radius, without resorting to the scaling laws.  The resulting curves are  shown in the inset of Fig.~\ref{fig:V_AO}. Including the exact cluster size distribution and the exact behaviour of the gyration radius makes it possible to properly capture even the region $r< 2R_{s=50}$ with the simple superposition of the AO contributions. Hence, we conclude that the failure of the model for $r < 2R_{s=50}$ is not related to the $AO$ approximation, but to the non-asymptotic  (model-dependent) behaviour, which is clearly visible in the size dependence of $R_g$ for small clusters in Fig.~\ref{fig:data}.


\section{Discussion}  

We have demonstrated that two colloids in  a  gel-forming solution 
 experience an attractive effective potential, which becomes increasingly long-ranged on approaching the percolation transition. The range  of the effective interaction is controlled by the connectivity length of the sol and
diverges at the percolation transition. Such effective interaction originates from the confinement of the clusters size fluctuations between the colloids surface, thus providing a new Casimir-like effect driven by the clustering properties of the sol. These results extend the analogy between the percolation transition and a second-order critical point~\cite{Coniglio} to the context of effective interactions.
In the case of critical Casimir forces,  the long-range attraction arises from the confinement of the order parameter fluctuations in between the colloids. When the latter are located at distances smaller than the correlation length, large-scale fluctuations are not allowed to occur between the colloids along the $r$ direction, giving rise to a non-zero net force. Approaching percolation, it is  the cluster size  distribution that  becomes wider and larger and larger clusters appear.  Similarly to critical fluctuations, clusters whose diameter is larger than the colloids surface-to-surface distance are  excluded.

The use of a simple theoretical description in which  clusters are treated as non-interacting spheres has shown that the mechanism controlling the effective interactions can be assimilated to a depletion effect. 
We expect that the residual interaction among monomers of different clusters (that are not included in our simulations)  would not affect significantly our results. In fact in the solvent cluster phase, most of the particle-particle interaction is already accounted in the formation of clusters and  the only remaining relevant cluster-cluster contribution is related to excluded volume interactions.
The fractal nature of the percolation clusters, which favours their interpenetration,  and the small overall packing fraction of the sol  help in modelling the resulting depletion potential with a theory that neglects cluster-cluster interactions (in analogy with the standard polymer depletants for which the AO model was conceived).   It is interesting to note that the depletion mechanism has been invoked as a guiding analogy for interpreting
critical Casimir forces~\cite{Buzzaccaro,ParolaJCP}, where the increase of the correlation length of the critical domains has been regarded as an increase of the size of the depleting objects.
We finally stress that the current analysis is based on a two-body description of the effective potential. In analogy with critical Casimir forces~\cite{Matos}, we expect that many-body effects will become relevant close to  percolation when the interaction range becomes comparable to the colloid size.

Exploiting the percolation transition for generating long-range effective forces opens up a new way to use self-assembly properties of the solvent (or co-solutes) for controlling 
interactions between colloidal particles. Differently to the case of critical Casimir forces
that require tuning of the solvent properties close to one specific point (the critical point),
percolation can be achieved for a variety of sol-densities.

Finally, we have  shown that the effective potential is long ranged only in the case of chemical sols,
when bonds between monomers have an infinite lifetime. Experimental investigation of  such phenomenon 
requires thus a chemical sol close to gelation.
 However, we speculate that the same results hold also 
 in the case of reversible bonds but only in the limit in which bond lifetimes are
  significantly longer than the experimental time scales (i.e. experiments that probe
  only a single sol microstate). 
In this perspective, with the increasing availability of self-assembling particles with specific design of interactions~\cite{GlotzerSolomon,cossms} 
such as gel-forming DNA tetramers~\cite{preprint}, we expect that the realisation of these long-range effective forces could be implemented, paving the way for further manipulation of colloidal phase behaviour and dynamical arrest.


\section{Methods:}

\begin{figure*}[ht!]
\includegraphics[width=17cm]{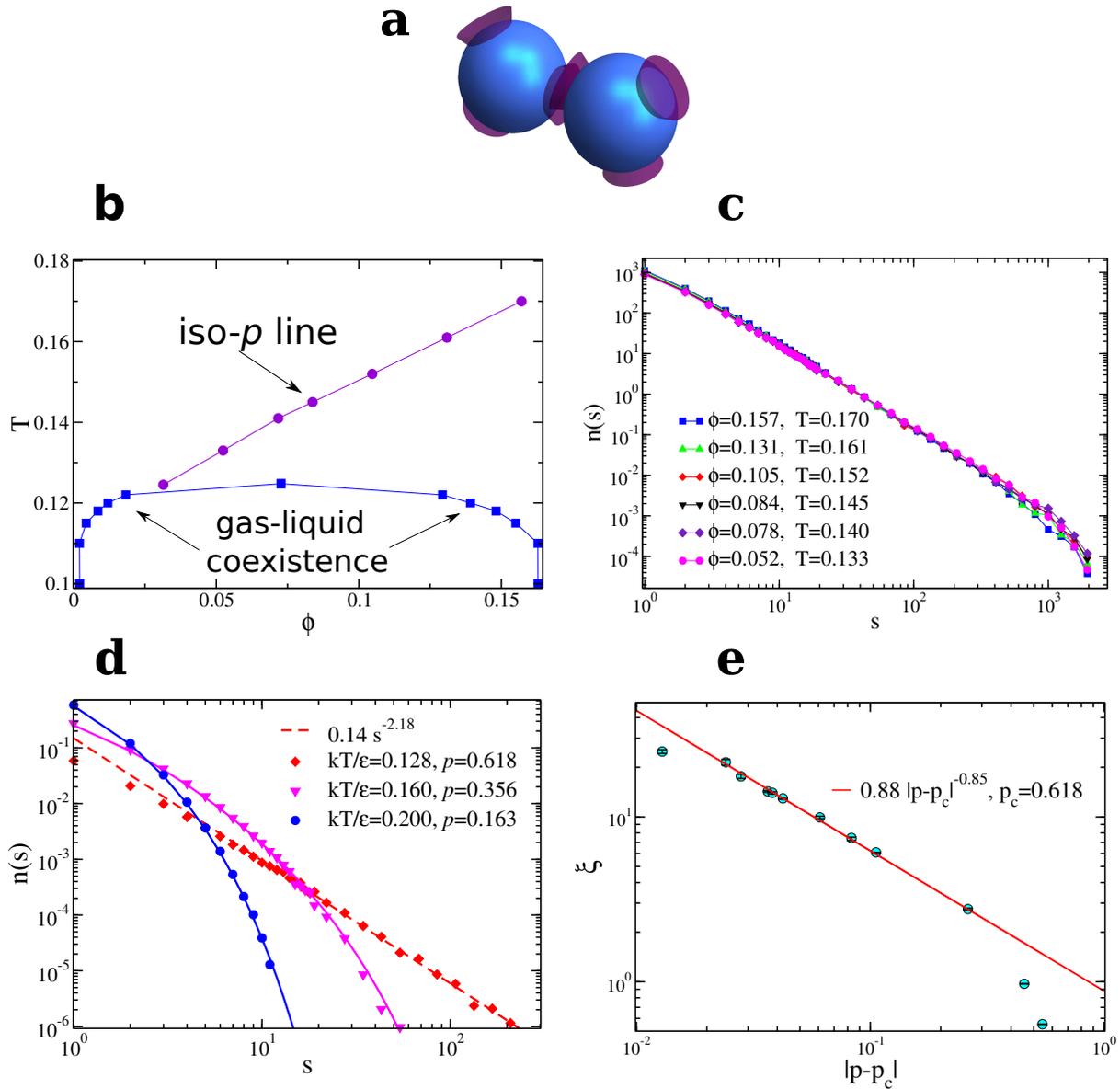}
\caption{\textbf{Clustering properties of the 3P particle model}  a) Sketch of two interacting 3P particles. 
 b) Phase diagram of the 3-patch particles model showing the gas-liquid (blue squares) coexistence line and a locus of  points (violet circles) of constant bond probability ($p \sim 0.565$).  c)   Cluster size distribution $n(s)$  for different densities along the $p \sim 0.565\pm 0.010$  locus. $n(s)$ is the same within the numerical precision.
d) Cluster size distribution of the Kern-Frenkel three-patchy model at $\phi=0.052$, for different temperatures (corresponding to different $p$). While at high $T$ ($p<<p_c$), $n(s)$ is well described by the 
mean-field Flory-Stockmayer distribution (solid lines), at the percolation point (corresponding roughly to $kT/\varepsilon=0.128$), $n(s)$ follows a power-law behaviour (dashed line) with exponent $\tau=2.18$.
e) Connectivity length $\xi$ of the sol at $\phi=0.052$ as a function of the distance from the percolation point. $\xi$ follows a power-law behaviour with the exponent consistent with the prediction of the random percolation theory ($\nu=0.88$). For $p\sim p_c$ finite size effects appears while for $p<<p_c$, the power-law behaviour breaks down. Here $p_c=0.618$. The standard error in evaluating the connectivity length over several configurations is also shown. }
\label{fig:3P}
\end{figure*}
\subsection{Model}
To generate the sol of clusters, we study a model of particles interacting  via the pairwise anisotropic Kern-Frenkel  three-patches (3P) potential~\cite{Kern}. Particles are thus represented by hard spheres of diameter $\sigma_m$ with three attractive sites, located on the equator. The interaction potential between these sites is

\begin{equation}
	V_{ij,\alpha\beta}=V_{\alpha\beta}^{SW}(|\vec r_{ij}|)G(\hat r_{ij},\hat r_{i\alpha},\hat r_{j\beta}),
\end{equation}
\noindent where $\vec r_{ij}$ is the vector between the centres of particles $i$ and $j$, and $\hat r_{i\alpha}$ is the unit vector from the center of particle $i$ to the center of the  $\alpha$ patch on the surface. $V_{\alpha\beta}^{SW}$ is a square well potential of width $\delta=0.119$ and depth $\varepsilon=1$
\begin{equation}
	V_{\alpha\beta}^{SW}(|\vec r_{ij}|)=\left\lbrace
 	\begin{array}{l}
 		\infty\quad \text{ if}\quad |\vec r_{ij}|<\sigma_m,\\
 		-\varepsilon \quad \text{ if}\quad \sigma_m\le |\vec r_{ij}|\le \sigma_m+\delta\sigma_m,\\	 
		 0  \quad \text{ otherwise}.
  	\end{array}
  	\right.
\end{equation}

\noindent The function $G$ modulates the potential and depends on the reciprocal orientation of two particles:
\begin{equation}
	G(\hat r_{ij},\hat r_{i\alpha},\hat r_{j\beta})=\left\lbrace
 	\begin{array}{l}

	   1 \text{ if}\quad\left\lbrace
		\begin{array}{l}
  	    	\hat r_{ij}\cdot\hat r_{i\alpha}>\cos(\theta_{max}),\\
      		-\hat r_{ij}\cdot\hat r_{j\beta}>\cos(\theta_{max}),\\
      	\end{array}
       \right.\\

	   0  \quad \text{ otherwise}.
  	\end{array}
  	\right.
\end{equation}

\noindent The angular width that controls the volume available for bonding, is set to $\cos(\theta_{max})=0.894717$. 
A sketch of the $3P$ model is reported in Fig.~\ref{fig:3P}(a).  The 3P system is a model for a physical gel, i.e. it forms reversible clusters and its connectivity properties depend on the temperature T and the packing fraction $\phi$ . Hence, in principle, it is possible to tune $T$ and  $\phi$ in order to find state points 
with the desired number of inter-monomers bonds.  The phase diagram of the model 
in the $T$-$\phi$ plane, characterized by the  limited-valence gas-liquid phase separation \cite{Bianchi}, is shown in Fig.~\ref{fig:3P}(b).   Lines of constant number of bonds  --- i.e. lines of equal  bond probability (iso-$p$ lines)  --- identify loci of similar sol structure.  One of these lines is  also shown in Fig.~\ref{fig:3P}(b) . Along 
this line, the cluster size distribution is found to be the same within numerical resolution (see Fig.~\ref{fig:3P}(c))  confirming that the monomers are aggregated in clusters of similar polydispersity.

\noindent 

In most of the work  we fix the packing fraction of the particles to $\phi=0.052$ and equilibrate the system at several temperatures, to  probe states with different fraction of bonds $p$ and hence different cluster size distribution. On cooling indeed the system forms larger and larger clusters, till a percolation point is reached where a spanning cluster appears.  
 
Note that $3P$ particles  form transient clusters due to the reversibility of the bonds.  
To generate a model for chemical gel, after equilibration has been reached, 
we freeze all the bonds formed in an arbitrary configuration of the system, thereby making the clusters lifetime infinite. Once all bonds are frozen, particles belonging to different clusters cannot form new bonds and clusters behave as translating and rotating rigid objects interacting only via excluded volume  among them and with the two HS colloids.  This polydisperse set of clusters is then used as a chemical sol model in the 
numerical study of the effective potential between two colloids.

\subsection{Identification of the percolation threshold}
For the system under study (with $f=3$ bonding sites for each of the $N$ particles) the fraction of bonds can be calculated from the relation $p=-2\langle U\rangle/Nf$, where $\langle U\rangle$ is the average potential energy and the factor two accounts for all the bonded sites of the system. 
At the percolation point $(T^*,\rho^*)$, $p$  reaches its critical value $p_c$.
To roughly identify the percolation threshold for the 3P system at the packing fraction $\phi=0.052$ we have performed MC simulations of the 3P fluid in the canonical ensemble for different temperatures $T$ (i.e. for different $p$) in the absence of the two large HS colloids. 
The percolation point can be identified by studying the evolution of the cluster size distribution $n(s)$ for different state points, which has the form of a power-law at  $p=p_c$, i.e. $n(s) \sim s^{-\tau}$ (with $\tau=2.18$), while as $p\rightarrow p_c$  $n(s)$ is controlled by a scaling function, typically assumed to be exponential, that modulates the power-law (see Eq.~\ref{eq:ns}). The result is shown in Fig.~\ref{fig:3P}(d). While at high temperatures (far from percolation)  the function $n(s)$ follows the mean-field Flory-Stockmayer theory~\cite{Stockmayer,Flory}, at $kT/\varepsilon=0.128$ it is described by a power-law behaviour with the random-percolation universality class exponent~\cite{Stauffer}. Hence, for a system of $N=10836$ 3P particles at packing fraction $\phi=0.052$, we locate to a good approximation the percolation point at $kT/\varepsilon=0.128$ corresponding to $p_c=0.618$.

Once $p_c$ is identified we can study the behaviour of the connectivity length $\xi$ defined in Eq.~\ref{eq:xi2}, on approaching the percolation point. Close enough to the transition, random percolation theory predicts~\cite{Stauffer} that $\xi$ follows a power-law behaviour with exponent $\nu=0.88$.
Data for $\xi(p)$ are shown in Fig.~\ref{fig:3P}(e). The  power-law  scaling is observed only for  $p/p_c >0.8$. For this
reason we have carried our investigation  close enough ($p/p_c>0.8$) to the transition to observe a genuine effect associated  to the incipient percolating behaviour.

\subsection{Numerical evaluation of the effective potential}
To evaluate the effective potential it is necessary to calculate $P(r)$, i.e. the probability that the two colloids are found at distance $r$. Hence we implement an 
Umbrella Sampling (US) scheme~\cite{GnanJCP,FrenkelBook} which allows for a convenient parallelization of the code and to optimize the computational time by "flattening" the energy barrier that can be created by the presence of large clusters and  that can prevent the colloids to sample uniformly all the distances.
To probe the whole distance range, we perform $40$ parallel runs in which the two colloids explore $40$ different windows. Hence in the US scheme employed, for each simulation the two colloids sample only a small window of distances $\Delta_i$. 
 Each run  is a standard  Monte Carlo (MC) simulation in the canonical ensemble, where
both colloids and clusters are allowed to move with a size-dependent  MC step allowing for a 30\% acceptance.
$V_{eff}$ is evaluated by constraining the two colloids to move in a window $\Delta_i$ along the $x$-axis  (identifying the $r$-direction) of a parallelepipedal box where the length of the $x$-edge  $L_x =7.6\sigma_c$ is twice the length of $L_y$ and $L_z$. This guarantees that, for all the simulated state points, the surface-to-surface distance between colloids (and their periodic images) in all directions is always larger than the distance at which $V_{eff}$ goes to zero. The cluster size distribution is identical in all the $40$ runs. During the single run we evaluate the  probability $P(r,\Delta_i)$ of finding the two colloids at a given distance r within the window $\Delta_i$. 

Then, the total probability $P(r)$ is obtained by merging together the $P(r,\Delta_i)$  resulting from all the parallel runs by means of a least-squares based algorithm. The effective potential is calculated from  the relation $\beta V_{eff}(r)=-ln(P(r))+C$  where C is a constant chosen imposing $V_{eff}(\infty)=0$.

 Each potential should be averaged over several different realizations of the cluster fluid. Due to the long computational time requested for evaluating $V_{eff}$ (roughly one month on 40 cores), our averaging is limited to  two different realizations of the cluster size for each distance from the percolation threshold. 
 
\subsection*{Acknowledgments}
NG and FS acknowledge support from ERC-$226207$-PATCHYCOLLOIDS. NG and EZ acknowledge support from MIUR-FIRB ANISOFT (RBFR125H0M). We thank C. Bechinger, S. Buzzaccaro, C. Maggi, A. Parola and R. Piazza for helpful discussions and suggestions.

\subsection*{Author contributions}
N.G. performed the simulations and the data analysis. All authors contributed equally 
to the design of the research, interpretation of the results and to the writing of the paper.

\subsection*{Competing financial interests}
The authors declare no competing financial interests.

\end{document}